# Robust Observer Based Methodology for Frequency and Rate of Change of Frequency Estimation in Power Systems

Abdul Saleem Mir, Abhinav Kumar Singh, *Member IEEE* and Nilanjan Senroy, *Senior Member, IEEE*

*Abstract*— **An observer based adaptive detection methodology (ADM) is proposed for estimating frequency and its rate of change (RoCoF) of the voltage and/or current measurements acquired from an instrument transformer. With guaranteed convergence and stability, the proposed methodology effectively neutralizes the effect of the measurement distortions like harmonics, decaying DC components and outliers by adding its counter negative. It is robust to noise statistics, performs well while encountering step changes in amplitude/phase and is demonstrably superior to its precursors as established by test results. A benchmark IEEE NETS/NYPS 16 machine 68 bus power system has been used for performance evaluation of robust ADM against its precursors and scaled laboratory setup based on OP5600 multiprocessors was used for establishing its real-time applicability.**

## I. Introduction

FREQUENCY and its rate of change are prime indicators of stability and serve as inputs for protection/control devices in a power system [1]-[2]. Therefore, swift and accurate estimation of frequency and RoCoF is a prerequisite to tackle potential power system instabilities and avert possible malfunctioning of protective devices and limiters [2]-[4]. However, large scale integration of renewable energy sources impair the estimation of these parameters due to their stochastic nature, inertia deficiency, and high ramp rates [1]-[5]. Estimation accuracy is further impaired by poor-fit signal models and non-ideal characteristics of the measurement system (instrument transformers and measurement chain) as it distorts the measured signals to some degree [6]-[7]. Inadequate quality of the estimated RoCoF has been identified as a cause behind false tripping of Loss of mains (LOM) relays as LOM relay uses estimated RoCoF as a protection metric [8]-[10]. Likewise, effectiveness of the fast frequency control (or protection) service relies on the accuracy of frequency and RoCoF estimates utilized in the derivation of the law/logic [11]-[14]. Stepped jump in the measurement signal parameters (amplitude/phase) can trigger or threaten to trigger cascaded control and protection system actions which can lead to generation loss and genuine problems in frequency management. For example, a generation loss of 1200MW following a major fire incident in southern California on August 10, 2016 was triggered by a stepped phase jump which caused the estimated frequency to cross the threshold [5]. It is therefore necessary to design a robust paradigm to estimate frequency and RoCoF to achieve desired levels of accuracy as specified by IEEE standards [15]-[17] for all possible measurement distortions and anomalies to enable timely action for events leading to possible disconnection of supply.

Using a direct derivative to obtain frequency/RoCoF is aggravated by the poor power quality waveforms obtained from the instrument transformers. In this context, various architectures for estimation of frequency and RoCoF have been reported ([18]-[22] and references therein). The frequency divider (intended for use in transient stability analysis) is an approximate method for estimation of power system frequency [23]. Network/machine impedances are required in its implementation. Methods like notch filters [19], discrete Fourier transform and its variants may not perform well during transient conditions [20]. Likewise, phase locked loops (PLLs) [20], Taylor-Kalman based estimation algorithm and related methods suffer from signal modelling inaccuracies [20]-[21]. Moreover their performance deteriorates due to fluctuations in the measurement and may not produce accurate estimates. However, its implementation requires prior information about signal model. An efficient rudimentary method of computing RoCoF from estimated frequency is the rolling window methodology. In this methodology RoCoF is computed by dividing the frequency deviation over the window length by time duration of the window. Averaging methods like this generally produce inaccurate estimates and their performance is significantly affected by the selection of window length [22]. Enhanced versions using Interpolated discrete Fourier transform (IDFT) such as generalized Taylor-weighted least squares based IDFT (GTWLS-IDFT) [21], IDFT based Kalman filter (IDFT-KF) [22] have been reported to be effective. However, their performance is affected by sudden amplitude and phase jumps, and decaying DC components (which are a consequence of line/load/generation switching, tap changing

This work was supported by MEITY Govt. of India and EPSRC UK under Grant EP/T021713/1. This is the authors' accepted version of TPWRS-01482-2020 (DOI: 10.1109/TPWRS.2021.3076562).

Abdul Saleem Mir and Abhinav Kumar Singh are with the School of Electronics and Computer Science, University of Southampton, UK (email: abdulsaleemmir@gmail.com ; a.k.singh@soton.ac.uk). Nilanjan Senroy is with the Department of Electrical Engineering, Indian Institute of Technology Delhi, India; (email: nsenroy@ee.iitd.ac.in).

of the transformers, stepped changes in power through inductive components (overhead line, transformer), and faults) [5], [10]. Moreover, their performance is affected by distortions (not accounted for in the signal model) and noisy measurements acquired from instrument transformers.

To address the aforementioned issues, a robust nonlinear observer based adaptive detection methodology (ADM) scheme is proposed. It ensures mitigation of heavy tailed impulsive noise, effective neutralization of the harmonics, outliers and DC components present in the measurements by adding its counter negative. It waives off the requirement of inverse of the transducer transfer function required to eliminate distortion [6]-[7]. It performs well even when measurement noise is quite high (3% to 15%). IEEE 16 machine 68 bus power system (Fig. 1, [24]) has been used for performance evaluation of the methodology against its contemporary methodologies and real-time tests (performed on a laboratory setup of OP5600 Opal-RT multiprocessors) of the case studies are presented to establish the applicability of the method. It is well suited for the realization of a range of control and protection applications [8]-[14], [25]-[29].

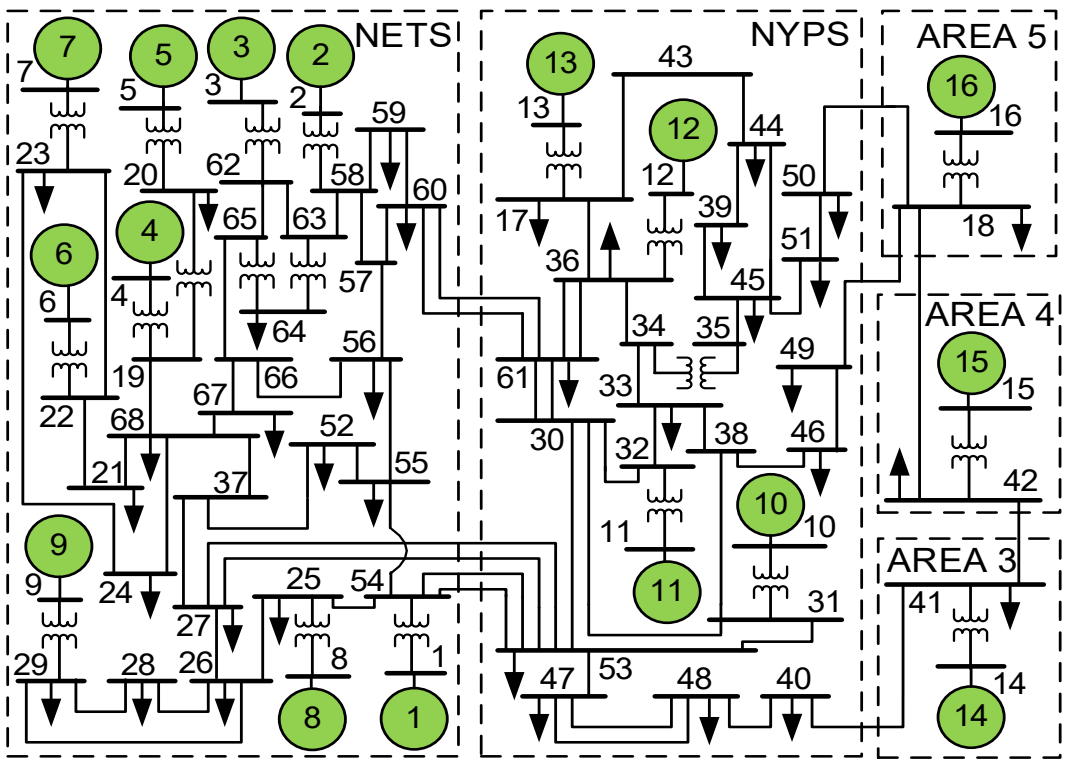

Fig. 1. IEEE NETS/NYPS 16 machine 68 bus system [24]

The contributions/advantages of this work are as follows:

- A Lyapunov criterion based robust ADM methodology has been derived and used to estimate the frequency and RoCoF from an analogue measurement without any prior knowledge of the measurement signal model. Assumed model of the signal accounts for distortions, harmonics and DC components and as such the ADM estimator is robust to measurement distortions.
- Robust ADM algorithm convergence is assured unlike currently available state of the art methodologies in the literature [18]-[22]. Convergence is corroborated by the *persistence-of-excitation* (PE) condition.
- The proposed methodology gives accurate estimates of frequency and RoCoF (with maximum percentage estimation errors less than 0.5%) when measurement noise (irrespective of its statistics) is high (2% to 15%).
- The ADM is robust to decaying DC outliers, stepped changes in the amplitude/phase and harmonic contamination in the measurements.
- The maximum and root mean square values of estimation errors (i.e., maximum, root-mean-square of the frequency error (FE) and maximum, root-mean-square of the RoCoF error (RE)) with robust ADM scheme are consistently and significantly lower than currently available methods in literature for all case studies [18], [21]-[22].
- Opal-RT multiprocessors based scaled laboratory setup has been used to validate the applicability of the developed methodology in *real-time* (RT). Actual field data was also used for performance evaluation.

The remainder of the paper is organized as follows: Instrumentation system and signal model has been discussed in section-II, the robust observer methodology (robust ADM algorithm) and its discrete implementation are discussed in section-III and section-IV respectively. Detailed case studies are presented in section-IV, real-time results in section-V whereas the conclusions are presented in section-VII.

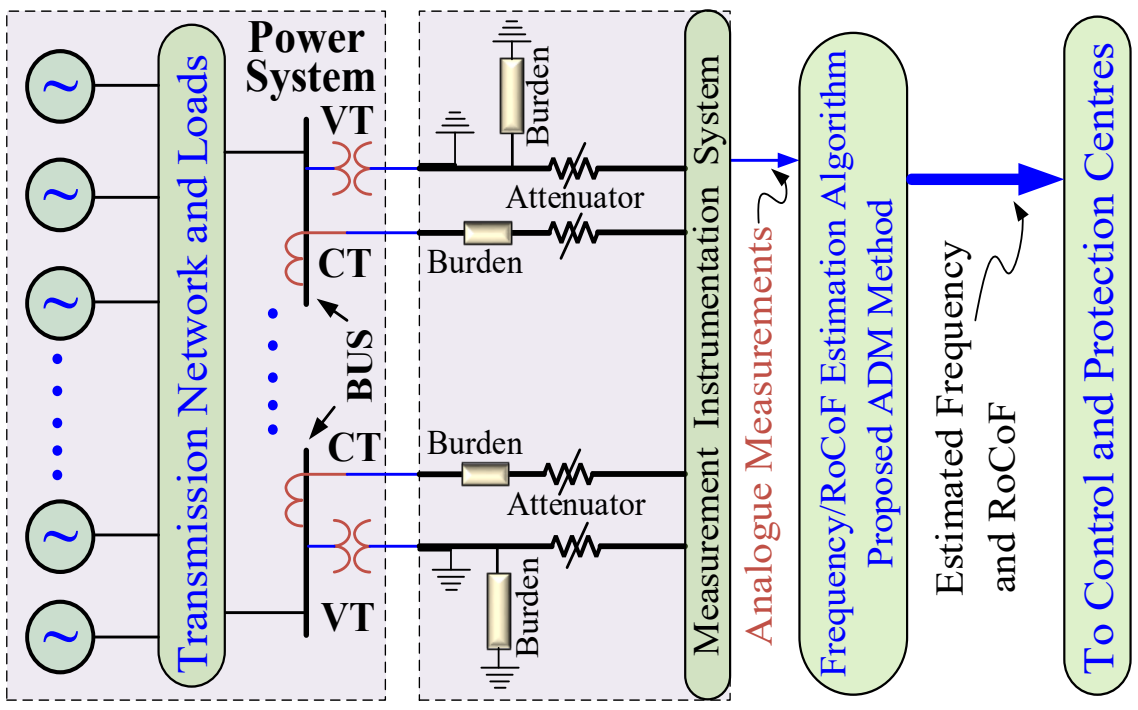

Fig. 2. Schematic of the frequency/RoCoF estimation methodology from analogue measurements.

## II. Field Instrumentation System and Signal Model

Errors in the analogue measurements for frequency and RoCoF estimation originate from an instrumentation system (which includes instrument transformers, instrumentation and control cables and associated burdens [30]) Fig. 2. In some cases instrument transformer (IT) saturates due to nonlinear characteristic of its core and it generates harmonics on its own [31]. The saturation of ITs, length of instrumentation cables and high burden resistance (Jonson noise [30], [32]-[34]) contribute to the errors and have an aggravating impact. Inaccurate frequency and RoCoF estimates stem due to application/usage of poor fit signal models as signal models used for frequency and RoCoF estimation do not account for aforementioned measurement distortions and anomalies and it can have potential practical implications [35]. Prior knowledge of the transducer transfer function inverse may be required to eliminate the distortion in the acquired measurements. The signal model (1) considered in this work is more generic and accounts for all the distortions. Signal model and robust ADM observer (discussed in next section) waives off the requirement of transducer transfer function inverse (required to eliminate the distortions [6]) from the estimation laws. Thus a generic measurement from an IT at a power system bus is represented mathematically as:

$$a(t) = \sum_{i=1}^{n} a_i(t) sin\big(\omega_i(t)t + \phi_i(t)\big) + a_{\mathrm{DC}} e^{-\frac{t}{\tau_{\mathrm{DC}}}} + \eta_\epsilon(t) \tag{1}$$

where, $a_i(t)$ is an amplitude, $\omega_i(t)$ the frequency in rad/s, $\phi_i(t)$ is a phase in rad, $a_{\mathrm{DC}}$ and $\tau_{\mathrm{DC}}$ are the parameters of DC component and $\eta_\epsilon$ accounts for noise and related distortions whereas $n$ is the harmonic order. The amplitudes, frequencies, phases and decaying DC component parameters ($a_{\mathrm{DC}}$ and $\tau_{\mathrm{DC}}$) are time evolving quantities.

## III. Robust Observer Methodology

The test system and the schematic of the methodology are shown in Fig. 1 and Fig. 2 respectively. Analogue measurements acquired from the instrumentation arrangement of the power system are processed by robust ADM scheme (detailed below) to output the frequency and RoCoF estimates. To derive the estimation laws, the measurement equation (1) is expanded as (2).

$$a = \sum_{i=1}^{n} a_{ci} sin(\omega_i t) + a_{si} cos(\omega_i t) + a_{\mathrm{DC}}\left(1 - \frac{t}{\tau_{\mathrm{DC}}}\right) + \eta_\epsilon$$

$$\Rightarrow a(t) = \Theta^{\mathrm{T}}\mathbb{S} + \eta_\epsilon \tag{2}$$

where,

$\mathbb{S} = [cos(\omega_1 t), sin(\omega_1 t) \ldots cos(\omega_n t), sin(\omega_n t), 1, (-t)]^{\mathrm{T}}$, $\Theta = [a_{s1}\ a_{c1}\ a_{s2}\ a_{c2} \ldots a_{sn}\ a_{cn}\ a_{\mathrm{DC}}\ a_{\mathrm{DC1}}]^{\mathrm{T}}$, $a_{si} = a_i sin\phi_i$, $a_{ci} = a_i cos\phi_i$, $a_{\mathrm{DC1}} = a_{\mathrm{DC}}\tau_{\mathrm{DC}}^{-1}$, $i = 1, \ldots, n$, $\omega_n = n\omega_1$. $\Theta$ is an unknown vector. Exponentially decaying DC term in the signal model has been approximated by first order Taylor's series expansion in (2). Higher order terms of the expansion are often significantly lower in magnitude. For parameter estimation the unknown vector '$\Theta$' is required to be estimated online. Hence,

$$\hat{a} = \widehat{\Theta}^{\mathrm{T}}\mathbb{S}\ ;\ \hat{a}_i = (\hat{a}_{si}^2 + \hat{a}_{ci}^2)^{1/2};\ \hat{\phi}_i = cos^{-1}(\hat{a}_{ci}/\hat{a}_i) \tag{3}$$

where, $\widehat{\Theta}^{\mathrm{T}} = [\hat{a}_{s1}\ \hat{a}_{c1}\ \hat{a}_{s2}\ \hat{a}_{c2} \ldots \hat{a}_{sn}\ \hat{a}_{cn}\ \hat{a}_{\mathrm{DC}}\ a_{\mathrm{DC1}}]$. $\hat{a}$ is the estimated measurement whereas $\hat{a}_i$ is the estimated value of $a_i$. The analogue measurement $a$ is observed through a transduction circuit with transfer function $\mathrm{G}(s)$.

$$\therefore \mathrm{z} = \mathrm{G}(s)\cdot a \Rightarrow \hat{\mathrm{z}} = \mathrm{G}(s)\cdot \hat{a} \tag{4}$$

$$\Rightarrow\ \tilde{\mathrm{z}} = \mathrm{G}(s)\cdot\tilde{a} = -\mathrm{G}(s)\widetilde{\Theta}^{\mathrm{T}}\mathbb{S} + \mathrm{G}(s)\eta_\epsilon \tag{5}$$

where, $\tilde{a} = a - \hat{a}$, $\tilde{\mathrm{z}} = \mathrm{z} - \hat{\mathrm{z}}$, and $\widetilde{\Theta} = \widehat{\Theta} - \Theta$. The transfer function $\mathrm{G}(s)$ is realizable and can be assumed as 'proper transfer function' [36]-[37]. Time-domain representation of the system ($\mathrm{G}(s) = \mathrm{C}(s\mathrm{I} - \mathrm{A})^{-1}\mathrm{B}$) is given by

$$\dot{\alpha} = \mathrm{A}\alpha + \mathrm{B}\left(-\mathbb{S}^{\mathrm{T}}\widetilde{\Theta} + \eta_\epsilon\right) = \mathrm{A}\alpha + \mathrm{B}\delta \tag{6}$$

$$\tilde{\mathrm{z}} = \mathrm{C}\alpha \tag{7}$$

where,

$\delta = -\mathbb{S}^{\mathrm{T}}\widetilde{\Theta} + \eta_\epsilon$ and $\alpha$ is an intermediate state vector. Equations (6-7) relate the parameter estimation error $\widetilde{\Theta}$ with $\tilde{\mathrm{z}}$ through $\alpha$. Therefore, Lyapunov function $\mathcal{L}(\widetilde{\Theta}, \alpha)$ (8) is constructed appropriately to derive the parameter update law.

$$\mathcal{L}\left(\widetilde{\Theta}, \alpha\right) = \tfrac{1}{2}\alpha^{\mathrm{T}}\mathcal{M}\alpha + \tfrac{1}{2}\widetilde{\Theta}^{\mathrm{T}}\Gamma^{-1}\widetilde{\Theta} \tag{8}$$

where, $\mathcal{M}$ is strictly positive and real (SPR) matrix and ensures positive definiteness of $\mathcal{L}(\widetilde{\Theta}, \alpha)$ and $\Gamma = \mathrm{diag}(\gamma_{c1}\ \gamma_{s1}\ \gamma_{c2}\ \gamma_{s2} \cdots \gamma_{cn}\ \gamma_{sn}\ \gamma_{\mathrm{DC}}\ \gamma_{\mathrm{DC1}})$ is a gain matrix with each element positive. Intermediate variable featuring in the Lyapunov function (8) cannot be measured/generated from (6) as $\delta$ is unknown. To waive off the explicit dependence of the parameter update laws upon intermediate state variable vector '$\alpha$', the lemma 1 has been used.

**Lemma 1** [36]-[37]: If the transfer function $\mathrm{G}(s)$ is real and positive, then there exists a vector $\boldsymbol{q}$, $\rho > 0$ and a matrix $\mathcal{M} = \mathcal{M}^{\mathrm{T}} > 0$ which satisfies the following expressions.

$$\mathcal{M}\mathrm{A} + \mathrm{A}^{\mathrm{T}}\mathcal{M} = -\rho\mathcal{N} - \boldsymbol{q}^{\mathbf{T}}\boldsymbol{q} \text{ and } \mathcal{M}\mathrm{B} = \mathrm{C}^{\mathrm{T}} \tag{9}$$

where, matrix $\mathcal{N} = \mathcal{N}^{\mathrm{T}} > 0$.

Using (9) the dependence of update law for estimation upon intermediate state vector $\alpha$ can be waived off. Therefore,

$$\dot{\mathcal{L}}\left(\widetilde{\Theta}, \alpha\right) = -0.5\alpha^{\mathrm{T}}\boldsymbol{q}^{\mathbf{T}}\boldsymbol{q}\alpha - 0.5\rho\alpha^{\mathrm{T}}\mathcal{N}\alpha - \tilde{\mathrm{z}}\delta + \widetilde{\Theta}^{\mathrm{T}}\Gamma^{-1}\dot{\widetilde{\Theta}} \tag{10}$$

The value of "$n$" in the signal model (1) is chosen to be very high such that the value of $\eta_\epsilon$ becomes significantly low. In that case, $\tilde{\mathrm{z}}\eta_\epsilon$ is extremely low. Therefore,

$$\dot{\mathcal{L}}\left(\widetilde{\Theta}, \alpha\right) \approx -0.5\alpha^{\mathrm{T}}\boldsymbol{q}^{\mathbf{T}}\boldsymbol{q}\alpha - 0.5\rho\alpha^{\mathrm{T}}\mathcal{N}\alpha - \tilde{\mathrm{z}}\widetilde{\Theta}^{\mathrm{T}}\mathbb{S} + \widetilde{\Theta}^{\mathrm{T}}\Gamma^{-1}\dot{\widetilde{\Theta}} \tag{11}$$

The parameter estimation law (12) ensures the negative semi-definiteness (NSD) of the time-derivative of the Lyapunov function candidate (13).

$$\dot{\widetilde{\Theta}} = \Gamma\tilde{\mathrm{z}}\mathbb{S}\ \Rightarrow \dot{\widehat{\Theta}} = \Gamma\tilde{\mathrm{z}}\mathbb{S} \tag{12}$$

$$\Rightarrow \dot{\mathcal{L}}\left(\widetilde{\Theta}, \alpha\right) = -0.5(\alpha^{\mathrm{T}}\boldsymbol{q}^{\mathbf{T}}\boldsymbol{q}\alpha) - 0.5\rho(\alpha^{\mathrm{T}}\mathcal{N}\alpha) \tag{13}$$

From (8-12), it can be deduced that $\mathcal{L}$, $\alpha$, $\tilde{\mathrm{z}}$, $\widetilde{\Theta}$, $\widehat{\Theta} \in \mathbf{L}_\infty$ [35],

$$\therefore \lim_{t\to\infty} \mathcal{L}\left(\widetilde{\Theta}\ (t), \alpha(t)\right) = \mathcal{L}_\infty < \infty \tag{14}$$

Equation (14) means that the Lyapunov function converges as it is a function of bounded variables, is positive definite and has a NSD time derivative (provided (13) holds). The individual update laws for the estimation of the signal parameters are given below as deduced directly from equation (12).

$$\dot{\hat{a}}_{c1} = \gamma_{c1}\tilde{\mathrm{z}}\ sin\ (\omega_1 t),\ \dot{\hat{a}}_{s1} = \gamma_{s1}\tilde{\mathrm{z}}\ cos\ (\omega_1 t),\ \cdots$$

$$\dot{\hat{a}}_{cn} = \gamma_{cn}\tilde{\mathrm{z}}\ sin\ (\omega_n t),\ \dot{\hat{a}}_{sn} = \gamma_{sn}\tilde{\mathrm{z}}\ cos\ (\omega_n t),\ \cdots$$

$$\dot{\hat{a}}_{\mathrm{DC}} = \gamma_{\mathrm{DC}}\tilde{\mathrm{z}}\ ,\ \dot{\hat{a}}_{\mathrm{DC1}} = -\gamma_{\mathrm{DC1}}\tilde{\mathrm{z}}t \tag{15}$$

where, $\gamma_{c1}\ \gamma_{s1}\ \gamma_{c2}\ \gamma_{s2} \cdots \gamma_{cn}\ \gamma_{sn}\ \gamma_{\mathrm{DC}}\ \gamma_{\mathrm{DC1}}$ are the estimation gains for parameters $\hat{a}_{c1}\ \hat{a}_{s1}\ \hat{a}_{c2}\ \hat{a}_{s2} \ldots \hat{a}_{cn}\ \hat{a}_{sn}\ \hat{a}_{\mathrm{DC}}\ a_{\mathrm{DC1}}$ respectively.

Vector "$\mathbb{S}$" should satisfy the *persistence-of-excitation* (PE) condition to ensure swift convergence of parameter estimates (Definition 4.3.1 in [36]). A piecewise continuous function $\mathbb{S}$: $\Re^+ \to \Re^n$ is persistently exciting in $\Re^n$ if there exist constants "$\rho_0$", "$\rho_1$", and "$\tau_x$" such that

$$\rho_0 I \le \mathrm{I}_{\mathrm{PE}} = \tau_x^{-1}\int_t^{t+\tau_x} \mathbb{S}(\tau)\mathbb{S}^{\mathrm{T}}(\tau)\,\mathrm{d}\tau\ \le \rho_1 I \tag{16}$$

Assuming the presence of decaying DC components and ignoring the permanent DC in the measurements, the PE property (16-17) can be proven. For measurements acquired from an instrument transformer, it is a realistic assumption as DC component ($a_{DC}(t) = a_{\mathrm{DC}}e^{-\tau_{\mathrm{DC}}^{-1}t}\ \forall\ t$ , $\tau_{\mathrm{DC}} > 0$) is a decaying exponential function.

As $t$ increases, $e^{-\tau_{\mathrm{DC}}^{-1}t} \to 0 \Rightarrow a_{DC}(t) \to 0$. The decaying exponential function ($e^{-\tau_{\mathrm{DC}}^{-1}t}$) has a stabilizing effect and the terms corresponding to this DC component in '$\mathbb{S}$' can be ignored in the derivation of PE condition.

In this context, "$\mathbb{S}\mathbb{S}^{\mathrm{T}}$" is a square matrix elements of which come from sinusoidal functions given below:

$$sin(\omega_i t)\ ,cos(\omega_i t),\ sin^2(\omega_i t),\ cos^2(\omega_i t),\ \cdots$$
$$sin(\omega_i t)cos(\omega_i t),\ cos(\omega_j t)sin(\omega_i t),\ sin(\omega_j t)sin(\omega_i t)\cdots$$
$$cos(\omega_j t)cos(\omega_i t),\ i,j \in [1,2,\dots,n] \text{ and } i \neq j.$$

With $\tau_x = 2\pi\omega_1^{-1}$, $\mathrm{I}_{\mathrm{PE}}$ in (15) reduces to:

$$\therefore \mathrm{I}_{\mathrm{PE}} = \begin{bmatrix} \pi\omega_1^{-1} & 0 & \cdots & 0 \\ 0 & \pi\omega_1^{-1} & \cdots & 0 \\ \vdots & \vdots & \ddots & \vdots \\ 0 & 0 & \cdots & \pi\omega_1^{-1} \end{bmatrix} \quad (17)$$

The ADM design methodology satisfies the PE condition (16) with $\rho_1 \geq \pi\omega_1^{-1}$ and $0 < \rho_0 \leq \pi\omega_1^{-1}$. Moreover "$\mathbb{S}$" and its derivative being vectors with sinusoidal elements are bounded i.e., $\dot{\mathbb{S}}, \mathbb{S} \in \mathbf{L}_\infty$. Satisfying these conditions (PE condition and $\mathbb{S}, \dot{\mathbb{S}} \in \mathbf{L}_\infty$), the exponential convergence of the estimates is guaranteed. Therefore,

$$\hat{a}_{c1} = \gamma_{c1}\mathrm{S}_\int = \gamma_{c1}\int_t\{\tilde{z}\, sin\,(\omega_1 t)\}dt \quad (18)$$
$$\hat{a}_{s1} = \gamma_{s1}\mathrm{C}_\int = \gamma_{s1}\int_t\{\tilde{z}\, cos\,(\omega_1 t)\}dt \quad (19)$$
$$\hat{a}_1 = \left(\gamma_{c1}^2\mathrm{S}_\int^2 + \gamma_{s1}^2\mathrm{C}_\int^2\right)^{1/2} \quad (20)$$
$$\hat{\phi}_1 = sin^{-1}(\hat{a}_{s1}/\hat{a}_1) \quad (21)$$

It should be noted that equations (11-12) ensure the convergence of other signal parameters (amplitude and phase). The Frequency estimation law is derived by minimizing the convex cost function (22) via steepest descent method (Appendix-B.2 in [36]). The estimated frequency (and RoCoF) converge to their respective true (actual) values when tracking error $\mathcal{E}$ goes to zero i.e., $\mathcal{E} \to 0 \Rightarrow \hat{\omega}_1 \to \omega_1^*$. $\omega_1^*$ is the actual value of $\omega_1$. A convenient way to ensure tracking error tends to zero is to adjust the frequency estimate in the direction that minimizes a quadratic cost function (22) of this tracking error via steepest gradient approach (23) (section 1.2.6 in [36]). The notion behind using square of the $\mathcal{E}$ (and not its absolute value) as objective function is to improve the accuracy and stabilization of the dynamic estimate trajectory.

$$\mathcal{J} = 0.5\mathcal{E}^2 = 0.5(\hat{a} - a)^2 \text{ where, } \mathcal{E} = (\hat{a} - a) \quad (22)$$
$$\therefore \dot{\hat{\omega}}_1 = -\eta'_\omega \partial\mathcal{J}/\partial\omega_1 \quad (23)$$
$$\Rightarrow \dot{\hat{\omega}}_1 = -\eta'_\omega \mathcal{E}\sum_{k=1}^{n}\{\hat{a}_{ck}ktcos(k\omega_1 t) - \hat{a}_{sk}ktsin(k\omega_1 t)\}$$
$$\approx -\eta_\omega \tilde{z}\sum_{k=1}^{n}\{\hat{a}_{ck}ktcos(k\omega_1 t) - \hat{a}_{sk}ktsin(k\omega_1 t)\} \quad (24)$$

where, $\eta'_\omega = |G(s)|\eta_\omega = |\tilde{z}/\varepsilon|\eta_\omega \Rightarrow \eta_\omega|\tilde{z}| = \eta'_\omega|\mathcal{E}|$ and $\eta_\omega$ is the tunable learning rate. Since $\eta_\omega$ and $\eta'_\omega$ are positive constants therefore $\eta_\omega\tilde{z} \approx \eta'_\omega\mathcal{E}$. Equations (15-16) (PE condition) and (11) ensure that the Lyapunov function (8) will have NSD time derivative. Therefore, the estimated parameters would eventually converge to their actual values for a given value of estimator gain $\Gamma$ which is tuned to ensure swift dynamic response of the parameter estimator. Standard particle swarm optimization algorithm (PSO) [38] was used to obtain the optimal value of the gain $\Gamma$ offline with *integral-square-error* (ISE) as cost function to ensure swift parameter convergence (Fig. 3) [38]. PSO generates a high-quality solution with stable convergence characteristic than other stochastic methods [39]. Likewise, it is a better option compared to empirical approaches as has been demonstrated in [38]-[39]. Traditional empirical methods lack the capability of forcing the estimate to follow actual time trajectory [39]. Therefore, ADM estimator gains were tuned optimally via PSO under different simulated scenarios (power system contingencies) to get an optimized value of the gain $\Gamma$.

## IV. Robust Observer: Discrete Time Realization

To realize the ADM algorithm (15) and (24) using a microcontroller or a digital signal processor (DSP), the estimate is calculated based on the previous value and sampled quantities as follows.

$$\hat{a}_{ci}^k = \hat{a}_{ci}^{k-1} + \mathrm{T_s}\gamma_{ci}\tilde{z}_k\sin(i\hat{\omega}_1^k(k-1)\mathrm{T_s}) \quad (25)$$
$$\hat{a}_{si}^k = \hat{a}_{si}^{k-1} + \mathrm{T_s}\gamma_{si}\tilde{z}_k\cos(i\hat{\omega}_1^k(k-1)\mathrm{T_s}) \quad (26)$$
$$\hat{a}_{\mathrm{DC}}^k = \hat{a}_{\mathrm{DC}}^{k-1} + \mathrm{T_s}\gamma_{\mathrm{DC}}\tilde{z}_k \quad (27)$$
$$\hat{a}_{\mathrm{DC1}}^k = \hat{a}_{\mathrm{DC1}}^{k-1} - (k-1)\mathrm{T_s^2}\gamma_{\mathrm{DC1}}\tilde{z}_k \quad (28)$$
$$\Delta\hat{\omega}_1^k = \hat{\omega}_1^{k+1} - \hat{\omega}_1^k \approx -\mathrm{T_s}\eta_\omega^k\partial\mathcal{J}_k/\partial\omega_1^k \quad (29)$$
$$\therefore \hat{\omega}_1^{k+1} - \hat{\omega}_1^k \approx -\mathrm{T_s}\eta_\omega^k\tilde{z}_k \times \quad (30)$$
$$\textstyle\sum_{i=1}^{n}\{\hat{a}_{ci}^k ik\mathrm{T_s}cos(ik\hat{\omega}_1^k\mathrm{T_s}) - \hat{a}_{si}^k ik\mathrm{T_s}sin(ik\hat{\omega}_1^k\mathrm{T_s})\}$$
$$\therefore \mathrm{RoCoF}_k = -(2\pi)^{-1}\eta_\omega^k\tilde{z}_k \times \quad (31)$$
$$\textstyle\sum_{i=1}^{n}\{\hat{a}_{ci}^k ik\mathrm{T_s}cos(ik\hat{\omega}_1^k\mathrm{T_s}) - \hat{a}_{si}^k ik\mathrm{T_s}sin(ik\hat{\omega}_1^k\mathrm{T_s})\}$$

and

$$\therefore \hat{f}_{k+1} = \hat{f}_k - \mathrm{T_s}(2\pi)^{-1}\eta_\omega^k\tilde{z}_k \times \quad (32)$$
$$\textstyle\sum_{i=1}^{n}\{\hat{a}_{ci}^k ik\mathrm{T_s}cos(ik\hat{\omega}_1^k\mathrm{T_s}) - \hat{a}_{si}^k ik\mathrm{T_s}sin(ik\hat{\omega}_1^k\mathrm{T_s})\}$$

where, $\mathrm{T}_s$ is sampling time. The laws (25-32) can be directly derived from the discrete equivalents of (8) and (22) given by equations (33) and (34) below respectively.

$$\mathcal{L}_k(\tilde{\Theta}_k, \gamma_k) = 0.5(\alpha_k^{\mathrm{T}}\mathcal{M}\alpha_k) + 0.5(\tilde{\Theta}_k^{\mathrm{T}}\Gamma^{-1}\tilde{\Theta}_k) \quad (33)$$
$$\mathcal{J}_k = 0.5\mathcal{E}_k^2 = 0.5(\hat{a}_k - a_k)^2 \quad (34)$$

Like "$\Gamma$", the choice of $\eta_\omega^k$ has to be appropriate. A smaller value of $\eta_\omega^k$ may guarantee convergence but at a very slow speed and estimate may not be acceptable. Too high $\eta_\omega^k$ is the cause of algorithm divergence.

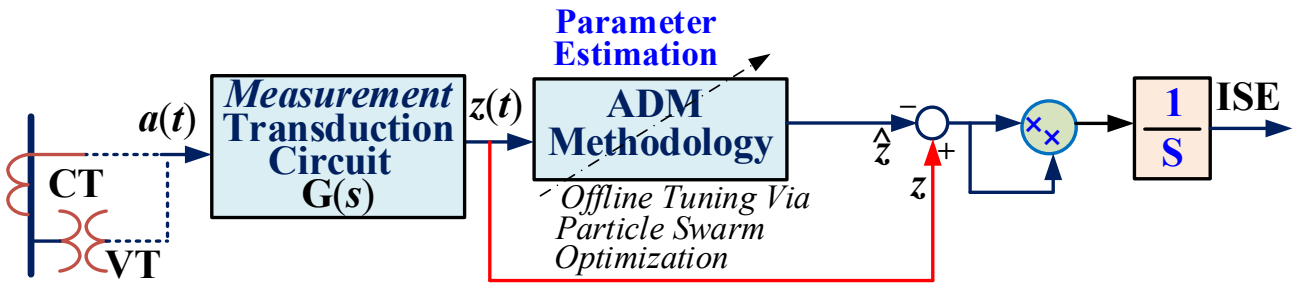

Fig.3: Schematic of the gain tuning of ADM algorithm via offline PSO with integral square error (ISE) as fitness function.

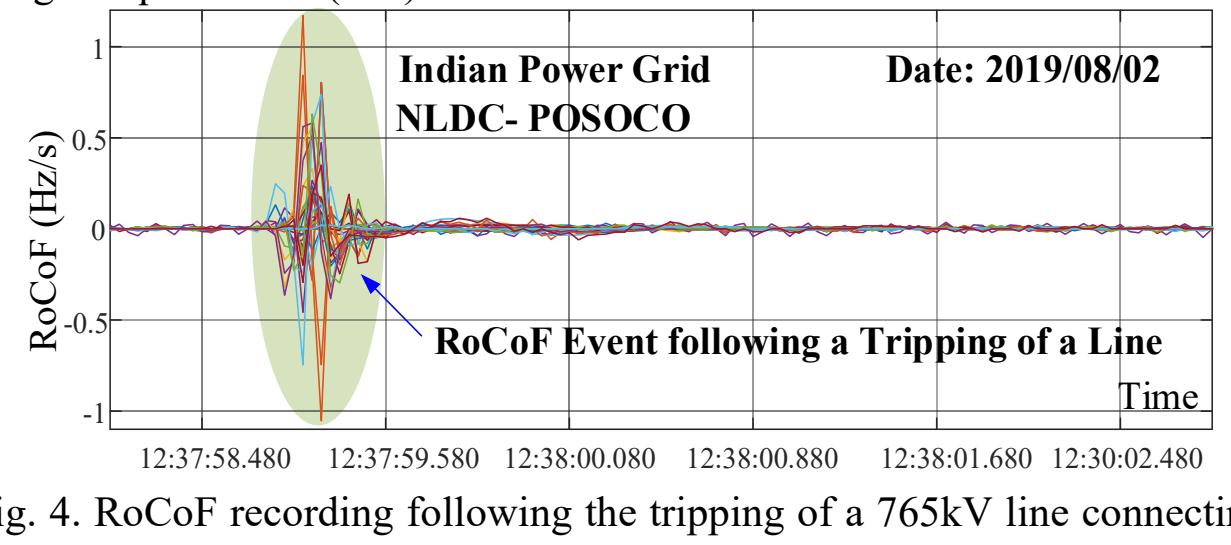

Fig. 4. RoCoF recording following the tripping of a 765kV line connecting Bilaspur and Rajnandangaon (Courtesy: NRLDC-POSOCO India).

TABLE I; ESTIMATION ERRORS : LATENCY= $100ms$, 2% NOISE

| Errors ↓ | PMU | GTWLS | IDFT-KF | ADM |
|---|---|---|---|---|
| Max (FE) ($Hz$) | 0.1865 | 0.0776 | 0.0675 | **0.0255** |
| RMSE (FE) ($Hz$) | 0.0584 | 0.0295 | 0.0142 | **0.0041** |
| Max (RE) (Hz/s) | 0.3776 | 0.2678 | 0.2311 | **0.1050** |
| RMSE (RE) (Hz/s) | 0.1879 | 0.0648 | 0.0493 | **0.0215** |

TABLE II; ESTIMATION ERRORS: LATENCY= $100ms$, 15% NOISE

| Errors ↓ | PMU | GTWLS | IDFT-KF | ADM |
|---|---|---|---|---|
| Max (FE) ($Hz$) | 0.7528 | 0.5212 | 0.4550 | **0.0882** |
| RMSE (FE) ($Hz$) | 0.1984 | 0.1295 | 0.0942 | **0.0241** |
| Max (RE) (Hz/s) | 0.6776 | 0.4678 | 0.4311 | **0.1551** |
| RMSE (RE) (Hz/s) | 0.3879 | 0.1648 | 0.1493 | **0.0615** |

**Theorem 1**: If the frequency/RoCoF is updated using (31-32), then estimated quantity tracks its actual time trajectory and the estimation error decays at an exponential speed provided adaptive gain $\eta_\omega^k$ satisfies (35) and $0 < \beta_\omega < 2$.

$$0 < \eta_\omega^{opt} = \frac{\beta_\omega \mathrm{T}_s^{-1}}{(\partial \mathcal{E}_k / \partial \widehat{\omega}_1^k)^2} < \frac{2\mathrm{T}_s^{-1}}{(\partial \mathcal{E}_k / \partial \widehat{\omega}_1^k)_{max}^2} \qquad (35)$$

**Proof**: With $\mathcal{J}_k$ as Lyapunov function, and using (34), the condition for $\eta_\omega^k = \eta_\omega^{opt}$ (35) can be proven (Appendix-A).

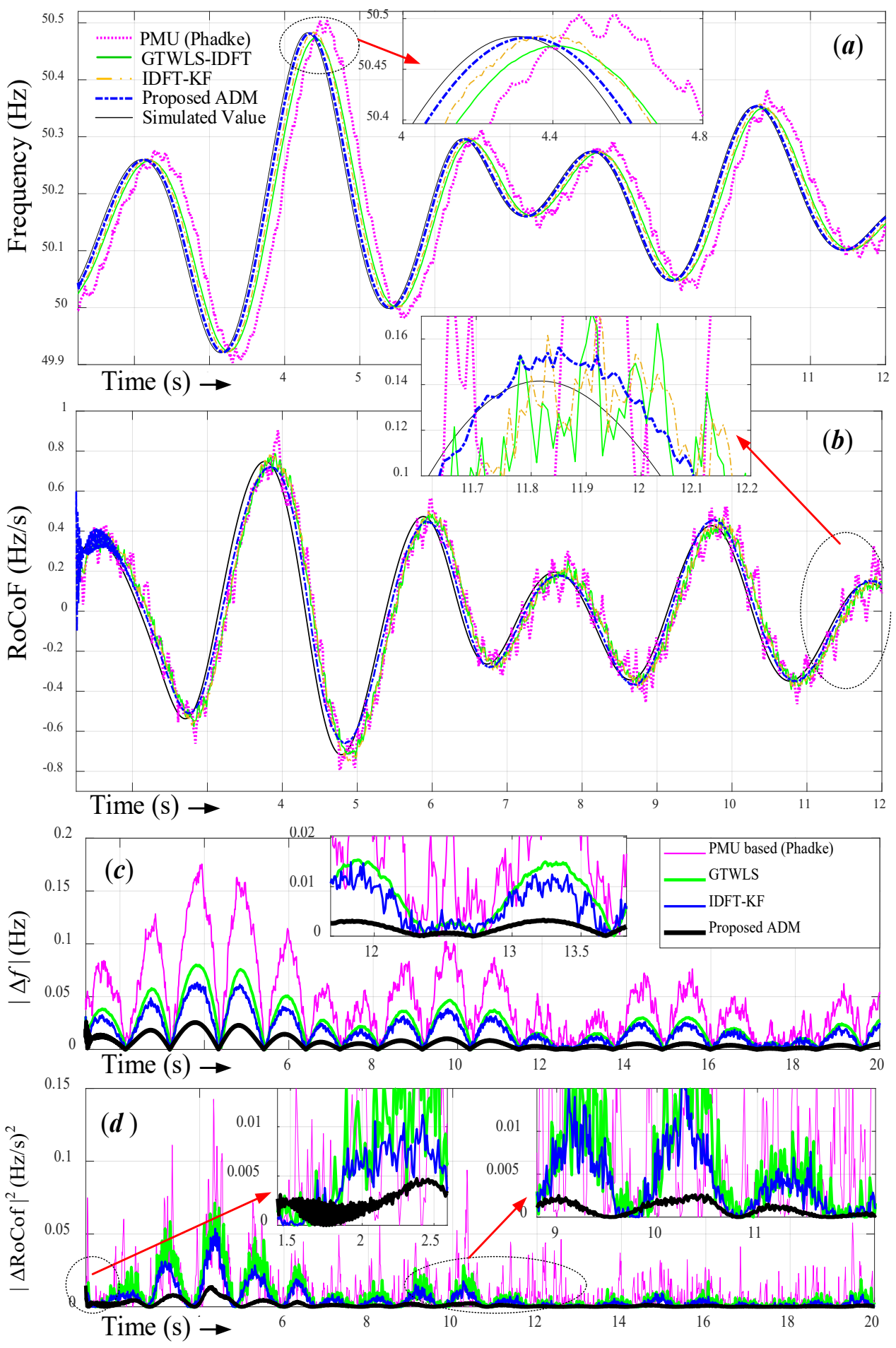

Fig. 5. Gaussian Case: (*a*) Estimated frequency from the analogue voltage at the terminals of the 13th generation unit of the test system (*b*) Estimated RoCoF from the analogue voltage at the terminals of the 13th generation unit of the test system (c) Performance comparison: Absolute value of the error in the estimated frequency (Hz) (*d*) Performance comparison: squared value of the error in the estimated RoCoF ($\mathrm{Hz}^2 s^{-2}$).

## V. CASE STUDY

The frequency and the RoCoF estimation via ADM was illustrated by subjecting the benchmark 16 machine 68 bus test system (Fig. 1) to a bolted 3-phase fault at bus 54 at $t = 1s$ and subsequently clearing the fault at $t = 1.18s$ by opening the faulted tie-line of the double circuit, to simulate RoCoF events as high as $1Hz/s$ (like the one shown in Fig.3 in an Indian power grid recorded at national load dispatch centre following a line switching event of 765kV Bilaspur-Rajnandangaon transmission line) and frequency deviation as high as $0.5\,Hz$ [15]-[16]. Case studies to test the ADM algorithm effectiveness are discussed as follows.

*a.* Case1: *Gaussian Noise*

For instrument transformers the measurement error range is given by $0.1\% - 3\%$ [40]-[41]. For this case study we have assumed an intermediate value of 2%. For performance evaluation the 'simulated values' of the frequency and the RoCoF were obtained directly from first and second derivative of the uncontaminated bus voltage (and/or current) phasor angles of simulated IEEE 16 machine 68 bus system. The simulated and the estimated frequency and RoCoF (for above disturbance scenario) of $\mathrm{V}_{13}$ (terminal bus voltage of 13th machine) are plotted in Fig. 5 (*a-b*). wherein the performance of the ADM method is compared to best known state of the art frequency and RoCoF estimation methods like PMU-algorithm [18], GTWLS-IDFT [21] and IDFT-KF [22]. The sampling frequency, window length and reporting rates of these algorithms are respectively chosen as 40 kHz, 1200 and 100 frames per second [22], [42]. For ADM the sampling rate was chosen as 24 samples per cycle (1.2 kHz). The corresponding estimation errors (absolute value of the frequency error (FE) in $Hz$ and square of the RoCoF error (RE) in ($Hz^2 s^{-2}$)) have been plotted in Fig. 5 (*c-d*). Moreover, the maximum and the root mean square errors (RMSE) of the FE and RE have been tabulated (Table-I). As inferred from the figure (Fig. 5) and the Table-I, the proposed ADM almost perfectly tracks the actual time trajectories of the frequency and RoCoF and performs significantly and consistently better than its precursors. Estimation latency of $100ms$ was selected for each method for performance comparison. To ascertain the comparative robustness of the ADM methodology the measurements were contaminated with noise levels as high as **15%**. Corresponding indices of frequency and RoCoF estimation accuracy were tabulated in Table-II. With ADM, the estimation accuracy indices (maximum and RMSE values of FE and RE) are significantly lower than the GTWLS and IDFT-KF. The maximum percentage estimation errors in the case of proposed ADM are less than 0.5%. The time instants at which there is a crossover of the time trajectories of the estimates with the actual trajectory of the frequency/RoCoF, the errors are zero. As soon as the estimation and the actual trajectories crossover, the estimation indices (absolute value of the frequency error (FE) in $Hz$ and square of the RoCoF error (RE) in ($Hz^2s^{-2}$)) increase first and then decrease till they approaches the next cross over point. Therefore trends in these estimation indices appear periodical. Similar periodical trends are observed in subsequent case studies Fig. 6-8.

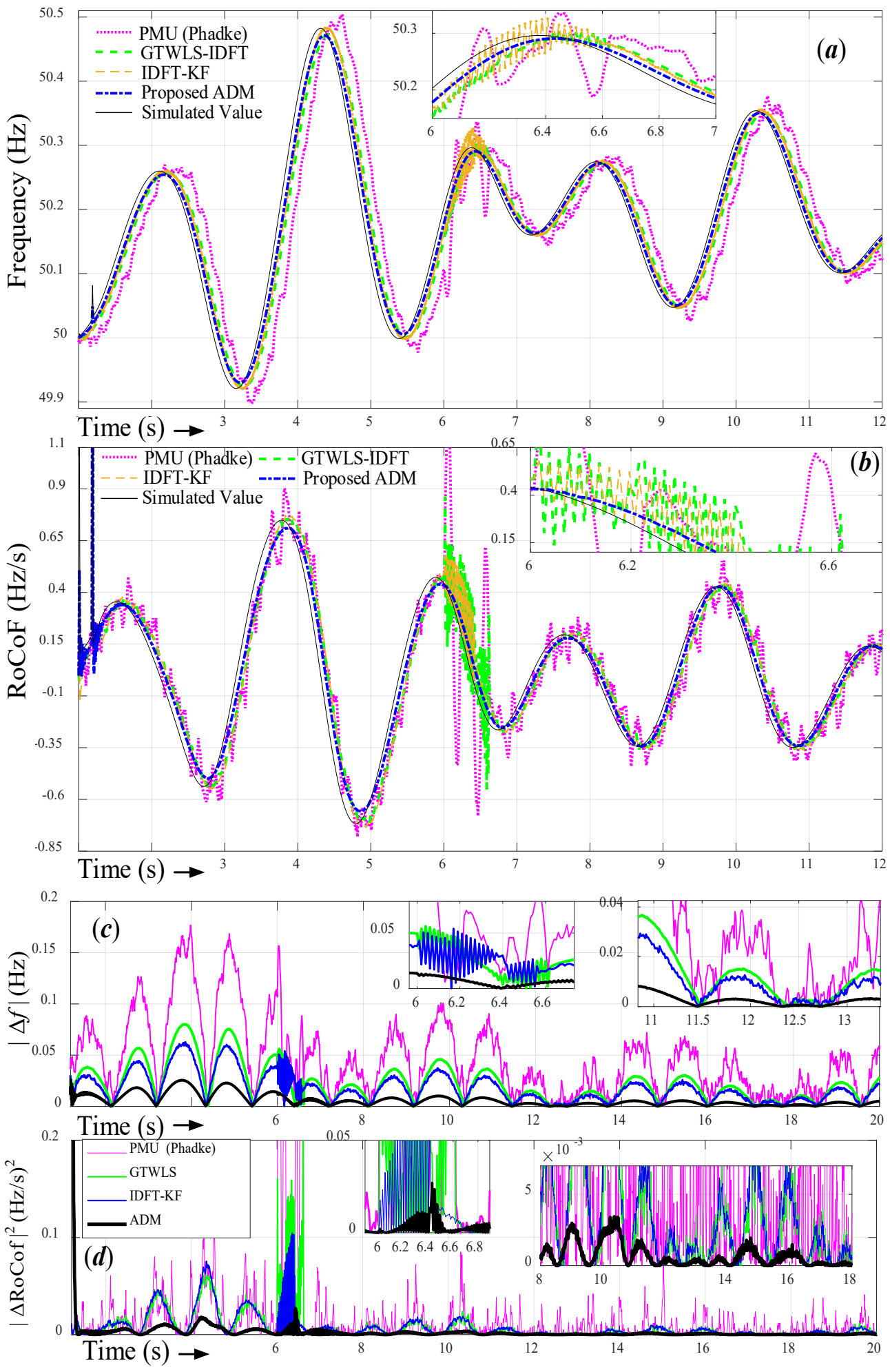

Fig. 6. Non-Gaussian Case: ($a$) Estimated frequency from the analogue voltage at the terminals of the 13th generation unit of the test system ($b$) Estimated RoCoF from the analogue voltage at the terminals of the 13th generation unit of the test system (c) Performance comparison: Absolute value of the error in the estimated frequency (Hz) ($d$) Performance comparison: squared value of the error in the estimated RoCoF ($Hz^2s^{-2}$).

## $b.$ Case2: *Non-Gaussian Noise and amplitude/phase steps*

For the same scenario as in Case 1, the measurements were assumed to be contaminated by 2% colored (non-Gaussian) noise in this case study. Additionally, an amplitude step of 0.05 p.u. (5%) and a small phase step of 0.04 radians were added to the measurements at $t = 6s$ for $0.4s$ to test the algorithm robustness against noise statistics and, stepped amplitude and phase changes. The corresponding estimation plots of frequency and RoCoF and errors for these quantities of the generator bus voltage $V_{13}$ of the 13th unit of the test system have been plotted in Fig. 6. Furthermore, as per the RoCoF test recommendations [15]-[17], the ADM algorithm and its precursors were tested exclusively for a large stepped phase change ($\pi/8$ radians) as well. The results have been plotted in Fig. 7. The choice of sufficiently large '$n$' [43] and addition of counter negative of measurement distortions enables the ADM to filter out biases and alleviate the effect of noise. The ADM estimator is therefore by design robust to noise statistics, changes in the amplitude/phase and large phase steps of the measured analogue signal and it performs better than its state of the art predecessors. Similar inferences may be drawn from the test results presented in Fig. 5-6 and table-I-II.

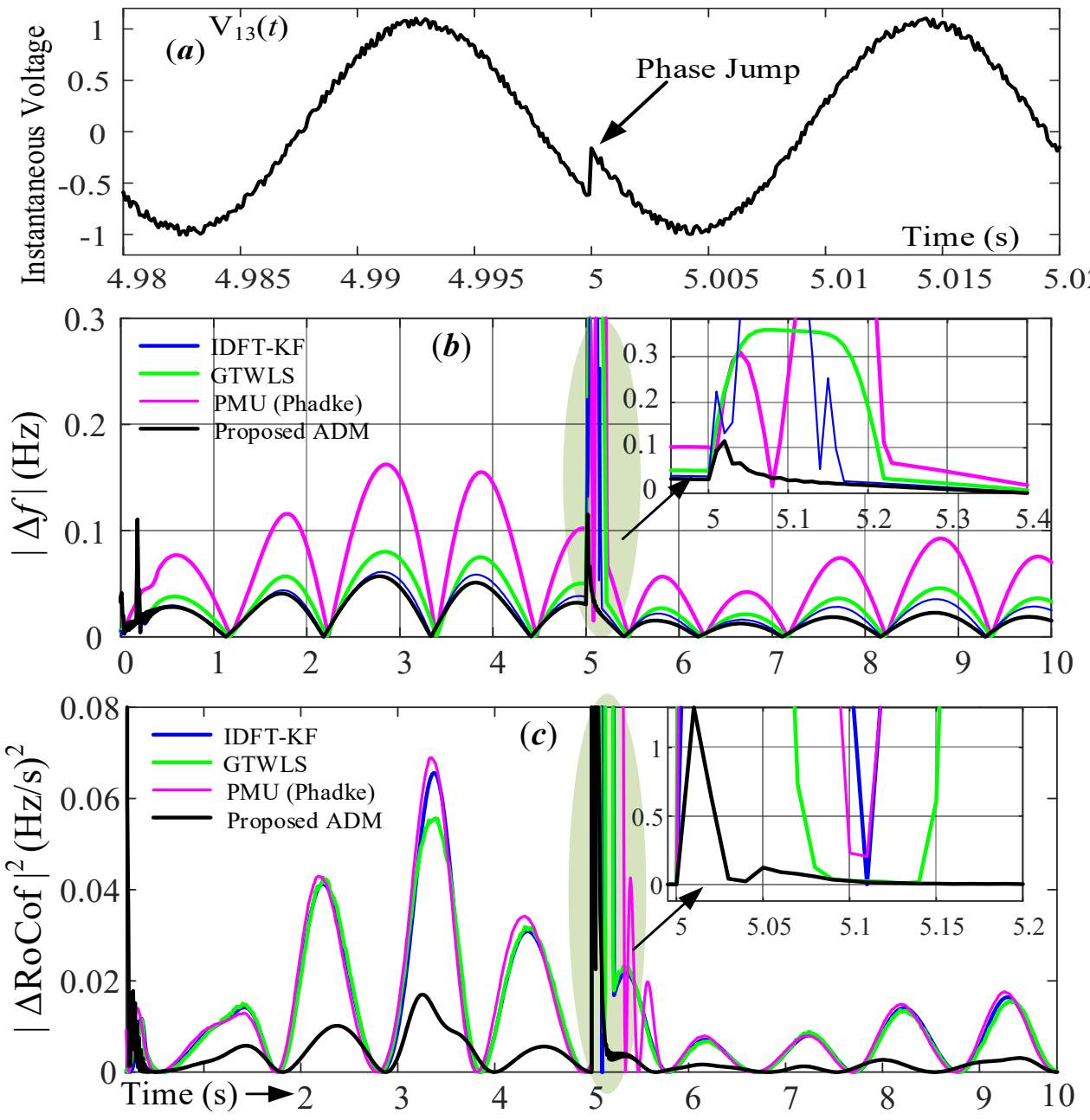

Fig. 7. Phase step in $\mathbf{V_{13}}$ of the 13th generation unit of the test system at $= 5s$ : ($a$) Phase step in analogue voltage ($b$) Performance comparison: Absolute value of the error in frequency estimation (Hz) ($b$) Performance comparison: squared value of the error in RoCoF estimation ($Hz^2s^{-2}$).

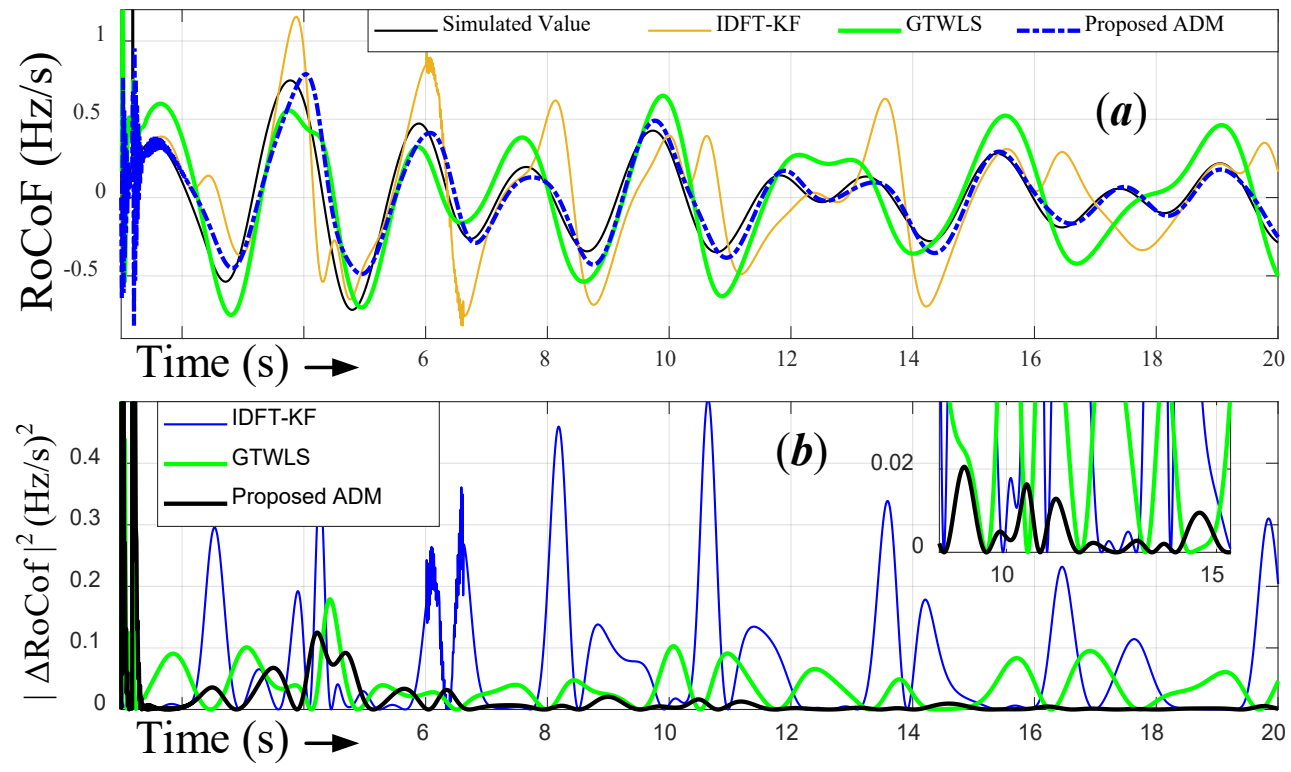

Fig. 8. Harmonic contamination Case: ($a$) Estimated RoCoF from the harmonically contaminated analogue voltage $\mathbf{V_{13}}$ ($b$) Performance comparison: squared value of the error in RoCoF estimation ($Hz^2s^{-2}$).

TABLE III; ESTIMATION ERRORS: LATENCY= $100ms$, 2% HARMONICS

| Errors ↓ | PMU | GTWLS | IDFT-KF | ADM |
|---|---|---|---|---|
| Max (RE) (Hz/s) | 1.1091 | 0.3519 | 0.7103 | **0.1642** |
| RMSE (RE) (Hz/s) | 0.3799 | 0.1021 | 0.2671 | **0.0874** |

TABLE IV; FIELD DATA TEST: RECONSTRUCTION ERRORS

| Reconstruction Errors → | PMU | GTWLS | IDFT-KF | ADM |
|---|---|---|---|---|
| | 1.000 | 0.3686 | 0.3263 | **0.2231** |

TABLE V; FIELD DATA TEST: ESTIMATED ROCOF ERRORS

| Maximum RoCoF Error (Hz/s) | | | |
|---|---|---|---|
| PMU Algo. [18] | GTWLS [21] | IDFT-KF [22] | **ADM** |
| 0.388 | 0.272 | 0.237 | **0.115** |

TABLE VI; SUBOPTIMAL LEARNING RATE AND ESTIMATION ERROR

| $\eta_\omega/\eta_\omega^{opt}$ | 1 | 1.02 | 1.04 | 1.06 |
|---|---|---|---|---|
| RMSE (FE) (Hz) | 0.0041 | 0.0049 | 0.0063 | 0.0073 |
| RMSE (RE) (Hz/s) | 0.0215 | 0.0299 | 0.0411 | 0.0481 |

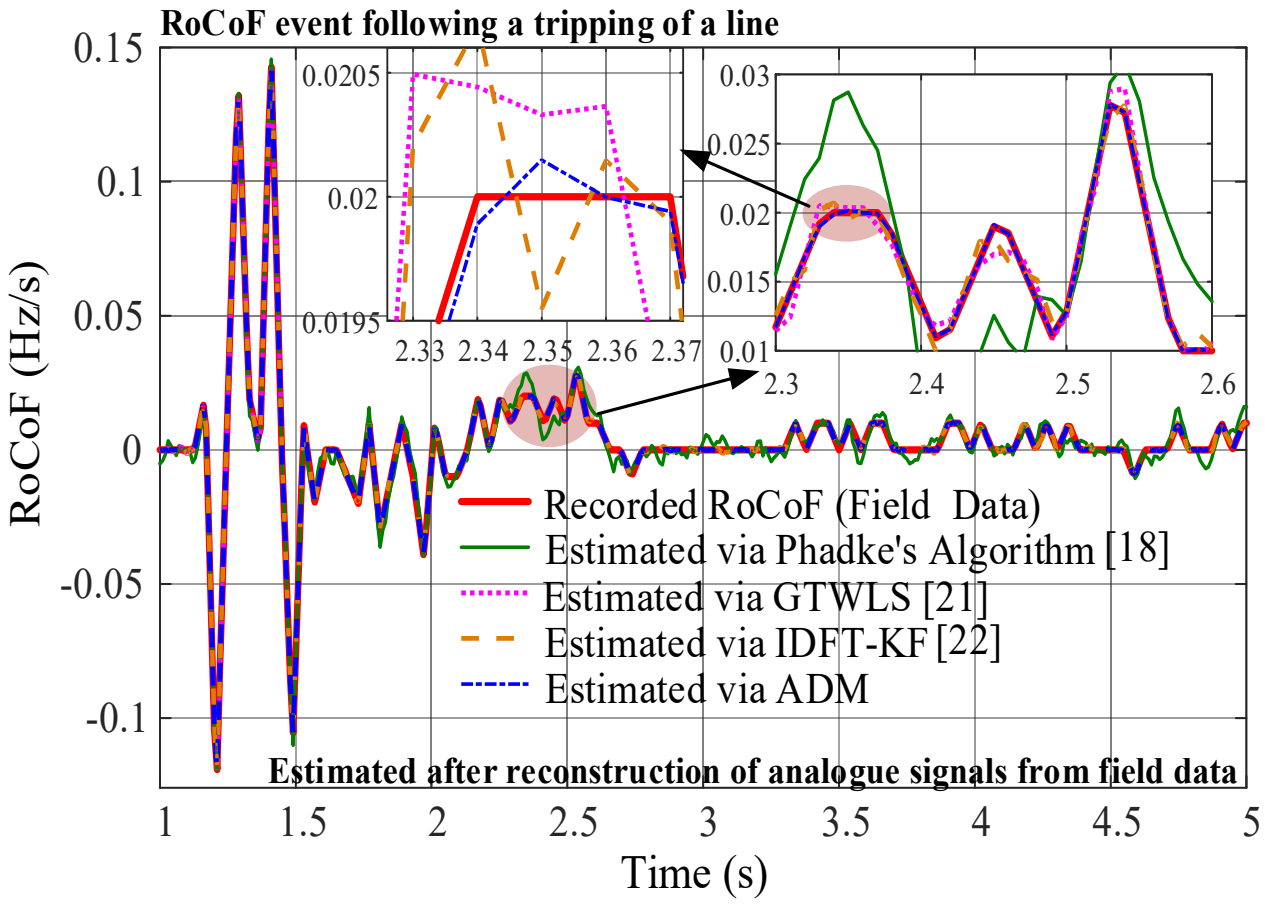

Fig. 9. RoCoF Estimation: Field data case study.

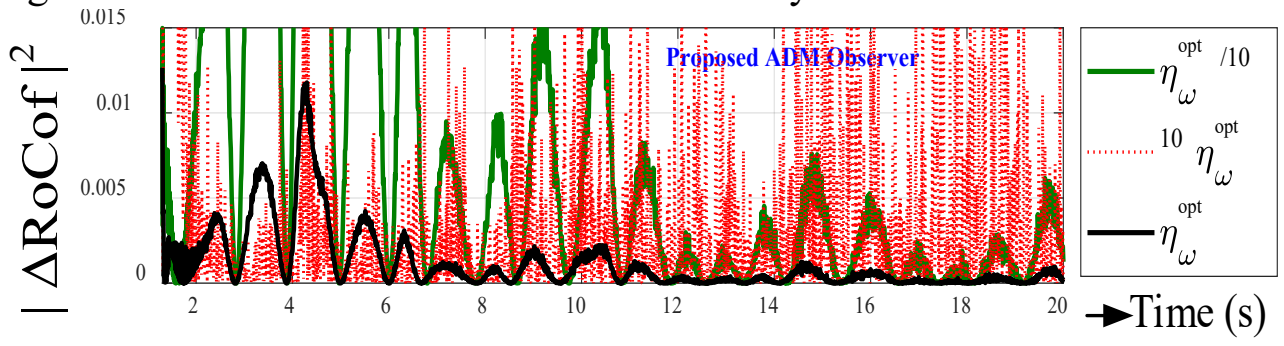

Fig. 10. ADM: RoCoF estimation errors and sensitivity to gain $\eta_\omega$.

### *c.* Case3: *Harmonics and Decaying DC Components*

For the same event as in Case 1, the input measurements were contaminated by 3rd harmonics (2%) throughout. Additionally, a decaying DC component was also added at t=1s in this case study. Most of the estimation methods fail due to signal modelling inaccuracies whereas ADM produces a good estimate of RoCoF as has been shown in Fig. 8. Corresponding accuracy indices of estimation showing effectiveness the scheme have been tabulated (Table-III).

### *d.* Case4: *Field Measurements*

To validate the accuracy of frequency and RoCoF estimation for an event or amalgam of events of significance in an actual power system, field synchrophasor measurements were used to construct the modulated analogue inputs (36) via lookup table and sampler for the estimation algorithm and subsequently contaminated by 2% noise and used to estimate the frequency and RoCoF. The outputs from the estimation algorithms were used to reconstruct the analogue waveforms and normalized reconstruction errors [43]-[44] were used as indices of performance. For **200** Monte Carlo runs, the normalized value of the indices have been tabulated (Table-IV). Furthermore, estimated and recorded RoCoF have been plotted in Fig. 9. whereas estimation index (maximum RoCoF error (MRE)) has been tabulated (Table-IV). ADM performs better than other methods (Fig. 9 , Table-IV-V).

$$a(t) = a_m sin\left(2\pi k T_s f + \pi k^2 T_s^2 \frac{df}{dt} + \theta_m\right) + \eta_\epsilon \quad (36)$$

where, $a_m$, $f$, $\frac{df}{dt}$, $\theta_m$, $T_s$, $k$ and $\eta_\epsilon$ are respectively the amplitude, frequency, RoCoF, phase, sampling time, sample number and noise of the phasor from the synchrophasor.

*Evaluation of design parameter $\eta_\omega$*

If the values of the design parameter $\eta_\omega$ is chosen either high or low than its optimal value ($\eta_\omega^{opt}$) described by (35), the estimation results are not good as has been shown in Fig. 10 for the same case study. It is therefore imperative to design this parameter optimally such that it doesn't violate the design condition (35) and yield the best performance. The optimal value $\eta_\omega^{opt}$ is obtained by performing exhaustive offline Monte-Carlo simulation runs at different buses of the power system for a range of disturbance scenarios using ISE as a performance index. For a given noise level, a suboptimal value of the learning rate (slightly higher than its optimal value) accelerates the swiftness/convergence of the estimation process. However, the corresponding root mean square error (RMSE) of the estimation also increases with the suboptimal value of the learning rate (Table-VI). While satisfying the design condition (35) and using $\eta_\omega^{opt}$ as an initial value, the learning rate is made self-tuning (37).

$$\eta_\omega^k = \beta_\omega \mathrm{T}_s^{-1}(\partial\mathcal{E}_k/\partial\hat{\omega}_1^k)^{-2} \quad (37)$$

However the variation in $\eta_\omega^k$ should be limited within 5% band (determined empirically) around $\eta_\omega^{opt}$ i.e., $0.95\eta_\omega^{opt} \leq \eta_\omega^k \leq 1.05\eta_\omega^{opt}$ for enhanced dynamic performance of the estimator and to have an appropriate tradeoff between swiftness and convergence of the estimates.

Moreover, the quality of the estimates and the convergence rate do not depend on the location of the estimator but on the frequency and noise in measurement, which do not vary significantly for any given bus in the system or with system size, as is clear from the final estimation law equations (25-32). Therefore, the location of estimator does not affect quality of the estimates.

### *e.* *Effect of Instrumentation Channel Parameter Variation*

Variation of parameters (saturation factor, length of the control cables, burden resistance etc.) in the measurement chain impart distortion and have an adverse impact on the quality of the measured data available for estimation [34]-[35]. To study the impact on the accuracy of the estimates, the saturation factor (SF) (Appendix-B) of the IT was varied over its range and its estimation accuracy indices were plotted (Fig. 11.). These indices remain constant for each scheme as long as SF is less than unity and increase continuously as SF is increased beyond unity. However, the impact is least in the case of an ADM based estimator (Fig. 11.). The impact of the variation in other parameters of the instrumentation chain is similar.

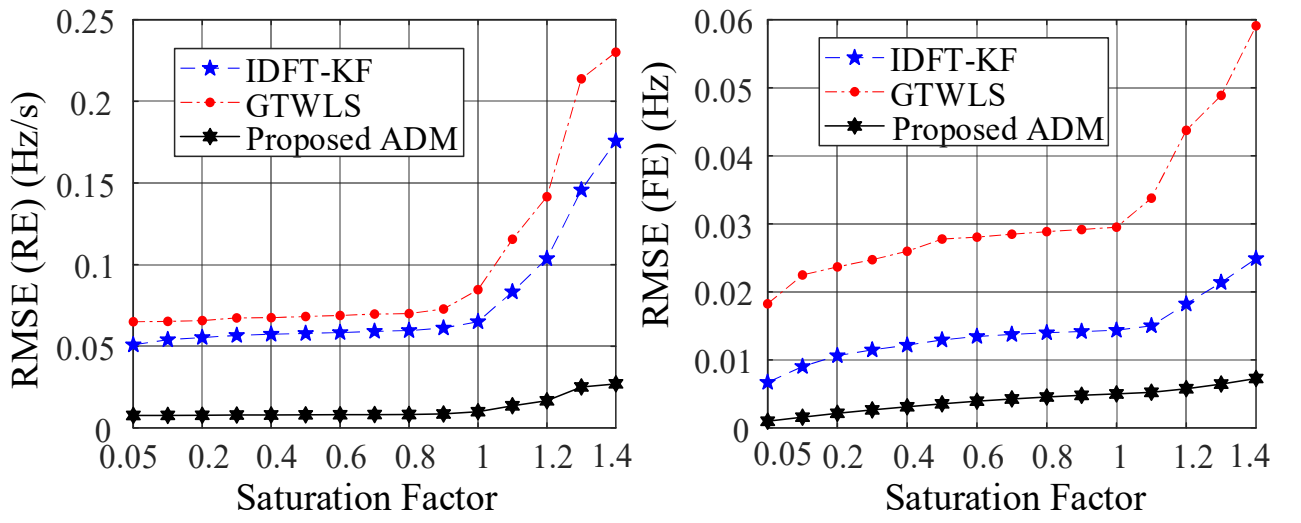

Fig. 11. Effect of instrumentation channel parameter variation (instrument transformer saturation factor) on estimation accuracy (*a*) Performance comparison: RMSE of the estimated RoCoF (Hz/s) with different algorithms. (*b*) Performance comparison: RMSE of the estimated frequency (Hz) with different algorithms.

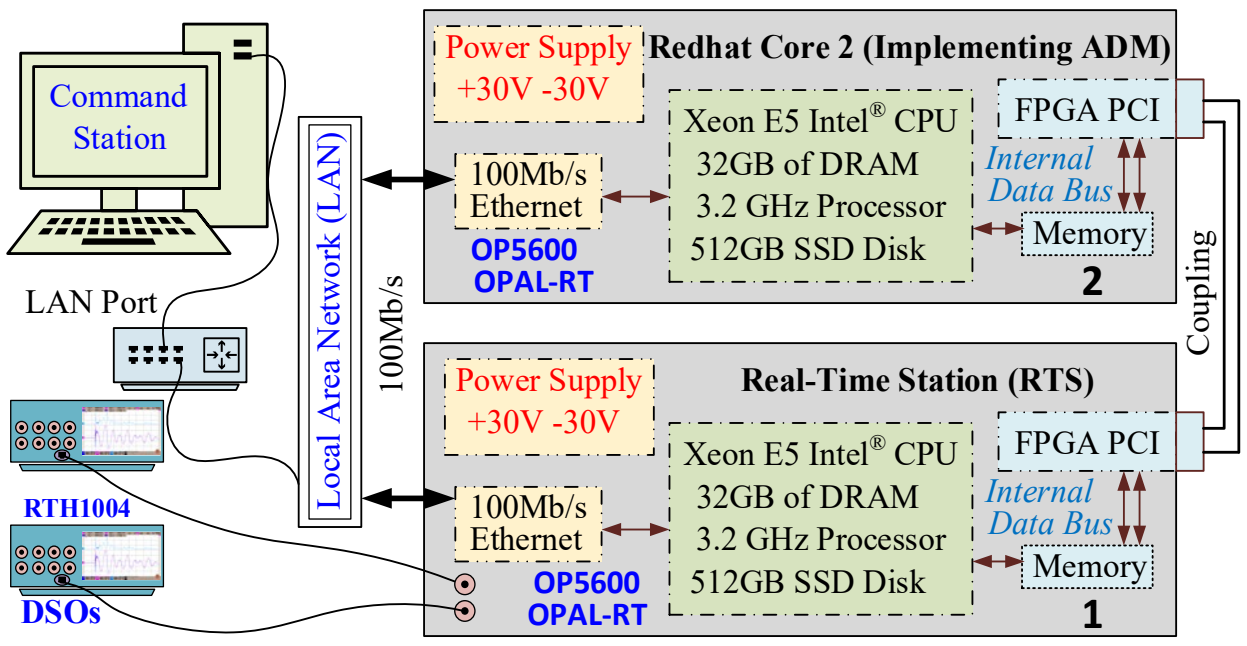

Fig. 12. Real-time implementation: Schematic Diagram.

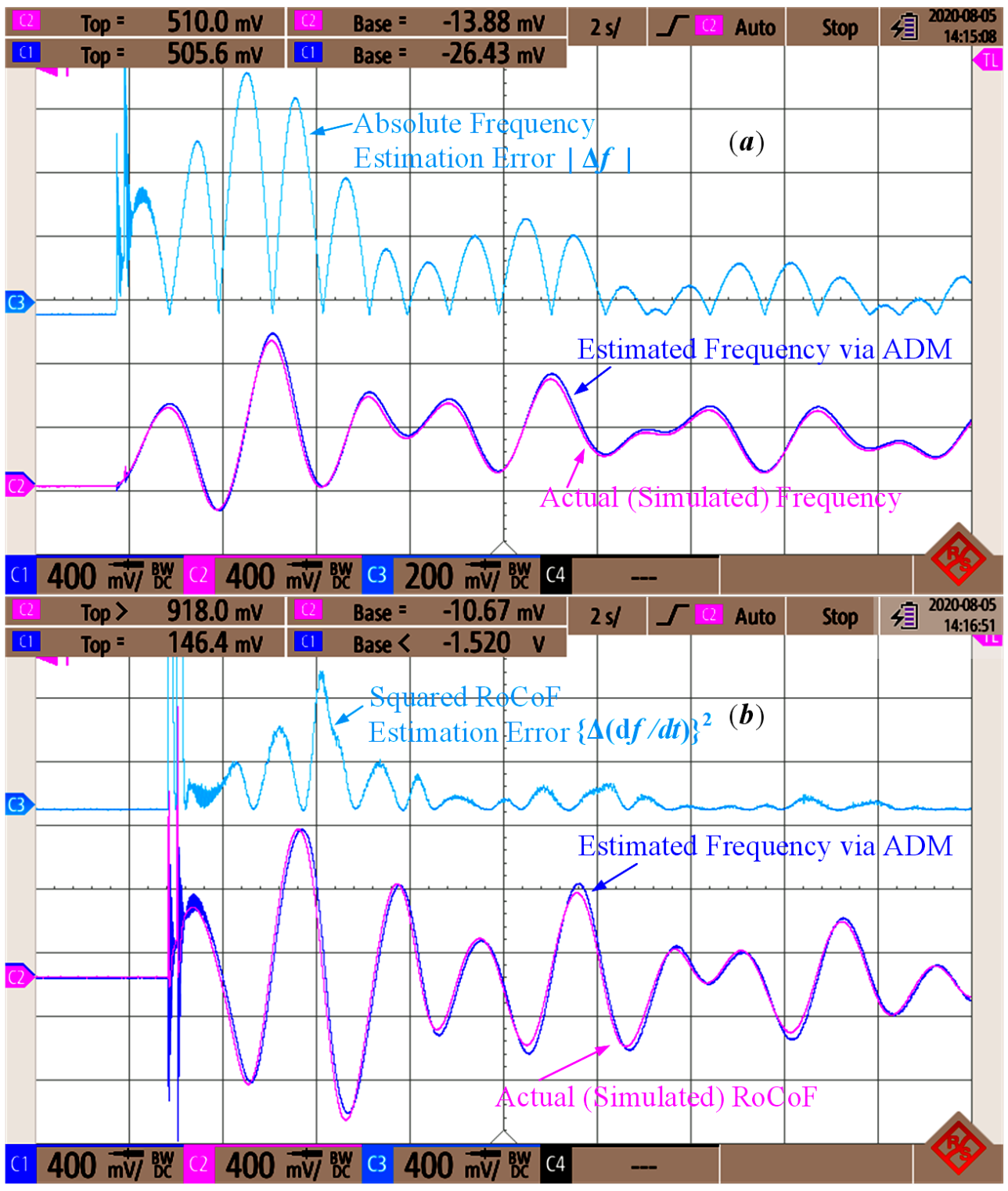

Fig. 13. Real-time results: Case1: Gaussian Case (*a*) Real-time frequency estimation via ADM (*b*). Real-time RoCoF estimation via ADM

## *f. Computational speed and viability*

The ADM takes an average running time of 0.21ms per iteration on a personal computer with $i$7-4970S CPU, 2GHz processor and 8GB RAM. The average computation time per iteration for IDFT-KF [22] and GTWLS [21] are 0.23ms and 0.19ms respectively. Therefore, the computational requirements of ADM can be easily met by an ordinary microcontroller/DSP processor.

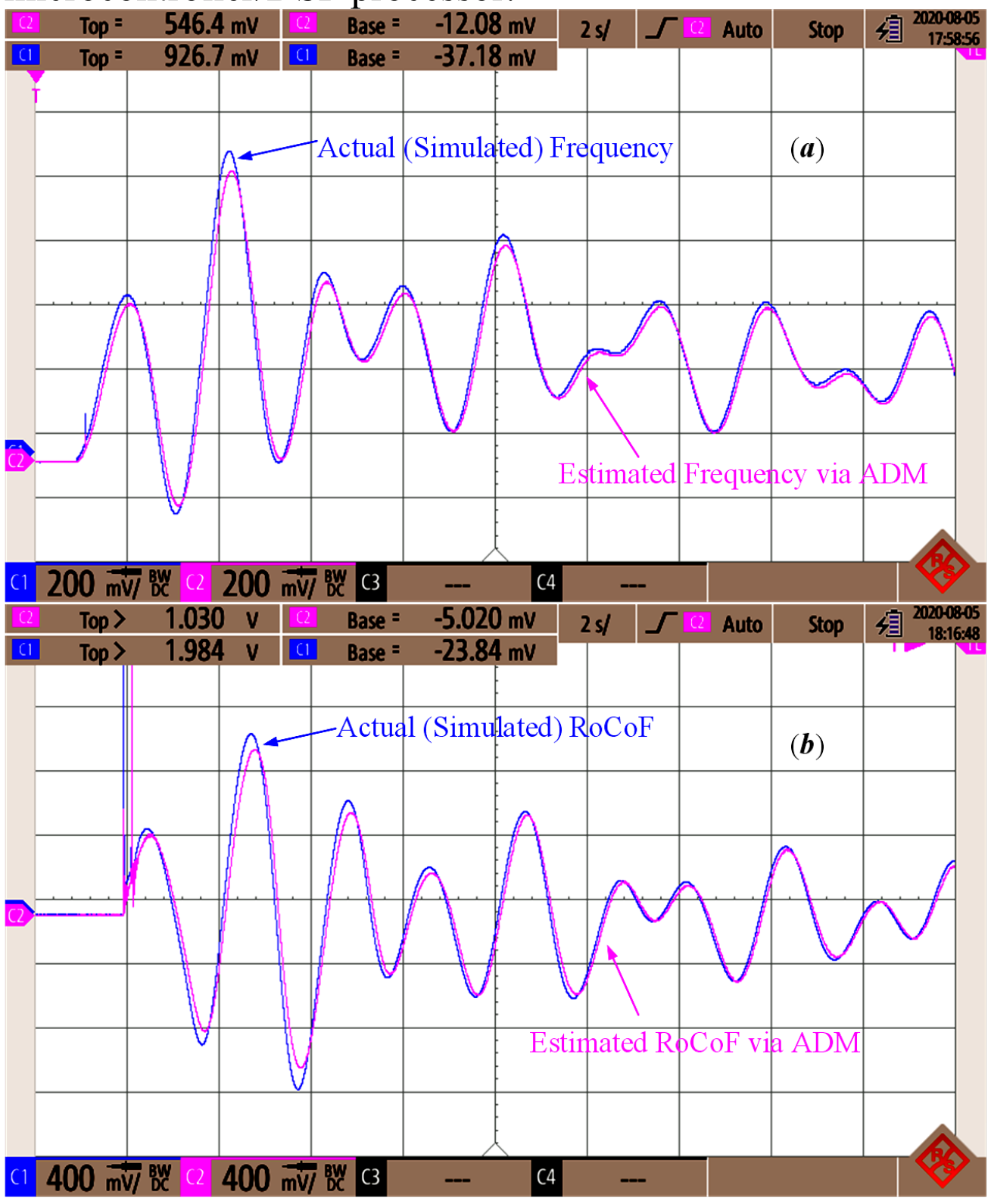

Fig. 14. Real-time results: Case2: non-Gaussian Case *a*) Real-time frequency estimation via ADM (*b*). Real-time RoCoF estimation via ADM.

## VI. Real-Time Implementation/Evaluation

A scaled laboratory setup has been used to demonstrate the realization and *real-time* applicability of the ADM scheme. Laboratory setup includes a windows-host command station (to set up test scenarios: Fig. 12.), QNX platform based Opal-RT multiprocessor as real-time station to emulate the Power System and network dynamics and a Linux Redhat OP5600 core implementing the ADM algorithm. The host command station, RTS and Redhat core2 (implementing ADM scheme) interact between themselves through 100Mbit/$s$ ethernet connection. The RT-lab main control is used to build real-time "**C**" code and to download and execute this code on Xeon processor of RTS through ethernet link. The same RT-lab main control is used to generate code for the ADM estimator and to download it to Redhat core2 for real-time execution through ethernet link. The interface between the two multiprocessor cores is in analogue domain through built-in DAC/ADC modules, so that it is virtually impossible for the ADM estimator to distinguish between the emulated plant and the actual plant. **RT1004** oscilloscopes were used to capture the test results of the executed scenarios detailed in section-V (*a-b*) as shown in Fig 13 (*a-b*) (wherein measurements are contaminated with 2% Gaussian noise) and Fig 14 (*a-b*) (wherein measurements are contaminated with 2% non-Gausssian noise). As inferred from the RT test case scenarios, ADM is robust and can be implemented using any processor with similar computational features as Op5600 Opal-RT processor.

## VII. Conclusion

In this paper, a new nonlinear observer (called robust ADM) based frequency and RoCoF estimator for power systems has been proposed to achieve functionalities like robustness to DC anomalies (stepped changes), distortions, harmonics and noise statistics (Gaussian or non-Gaussian). The ADM produces highly accurate estimates of frequency and RoCoF for all grid conditions. The proposed estimator is stable and the estimate convergence is guaranteed. It is therefore suitable for robust estimation. As established via theory and simulations, robust ADM performs better than state of the art methods in the literature owing to its self-adaptation to any changes in the measurement signal.

## Appendix A

**Proof**: With $\mathcal{J}_k$ as Lyapunov function, and using (34), the condition for $\eta_\omega^k = \eta_\omega^{opt}$ given by the equation (35) can be proven as detailed below [46].

$$\because \mathcal{J}_k = \mathcal{E}_k^2/2$$

Then, increment in $\mathcal{J}_k$ i.e., $\Delta\mathcal{J}_k$ is given by

$$\Delta\mathcal{J}_k = \mathcal{J}_{k+1} - \mathcal{J}_k = \frac{1}{2}[\mathcal{E}_{k+1}^2 - \mathcal{E}_k^2]$$

The error $\mathcal{E}_{k+1}$ is given by:

$$\mathcal{E}_{k+1} = \mathcal{E}_k + \Delta\mathcal{E}_k = \mathcal{E}_k + [\partial\mathcal{E}_k/\partial\widehat{\omega}_1^k]^T\Delta\widehat{\omega}_1^k$$

Substituting $\mathcal{E}_{k+1}$ in $\Delta\mathcal{J}_k$, we have

$$\Delta\mathcal{J}_k = \mathcal{E}_k\Delta\mathcal{E}_k + \frac{\Delta\mathcal{E}_k^2}{2}$$

$$\Rightarrow \Delta\mathcal{J}_k = \Delta\widehat{\omega}_1^k\left[\frac{\partial\mathcal{E}_k}{\partial\widehat{\omega}_1^k}\right]\left[\mathcal{E}_k + \frac{1}{2}\Delta\widehat{\omega}_1^k\left[\frac{\partial\mathcal{E}_k}{\partial\widehat{\omega}_1^k}\right]\right]$$

$\Delta\widehat{\omega}_1^k$ as obtained from (29-30) is given by:

$$\Delta\widehat{\omega}_1^k = -\eta_\omega^k \mathrm{T}_s \mathcal{E}_k[\partial\mathcal{E}_k/\partial\widehat{\omega}_1^k]$$

Substituting $\Delta\widehat{\omega}_1^k$ in $\Delta\mathcal{J}_k$.

$$\therefore\ \Delta\mathcal{J}_k = -\eta_\omega^k \mathrm{T}_s \mathcal{E}_k\left[\frac{\partial\mathcal{E}_k}{\partial\widehat{\omega}_1^k}\right]^2\left[\mathcal{E}_k + \frac{1}{2}\eta_\omega^k \mathrm{T}_s \mathcal{E}_k\left[\frac{\partial\mathcal{E}_k}{\partial\widehat{\omega}_1^k}\right]^2\right]$$

$$\Rightarrow \Delta\mathcal{J}_k = -\eta_\omega^k \mathrm{T}_s \mathcal{E}_k^2\left[\frac{\partial\mathcal{E}_k}{\partial\widehat{\omega}_1^k}\right]^2 + \frac{1}{2}(\eta_\omega^k)^2\mathrm{T}_s^2\mathcal{E}_k^2\left[\frac{\partial\mathcal{E}_k}{\partial\widehat{\omega}_1^k}\right]^4$$

$$\equiv -\varrho\mathcal{E}_k^2 \tag{A.1}$$

Let $h_{1k} = \frac{\partial\mathcal{E}_k}{\partial\widehat{\omega}_1^k}$ and $h_{max}^k = \left[\frac{\partial\mathcal{E}_k}{\partial\widehat{\omega}_1^k}\right]_{max}$, $\eta_\omega'' = \eta_\omega^k\,(h_{max}^k)^2$

$$\Rightarrow \varrho = \frac{1}{2}h_{1k}^2\eta_\omega^k(2 - \mathrm{T}_s\eta_\omega^k h_{1k}^2)$$

$$= \frac{1}{2}h_{1k}^2\eta_\omega^k\left(2 - \mathrm{T}_s\eta_\omega''\frac{h_{1k}^2}{(h_{max}^k)^2}\right)$$

Since, $h_{1k} < h_{max}^k$, therefore

$$\varrho \geq \frac{1}{2}h_{1k}^2\eta_\omega^k\mathrm{T}_s(2 - \mathrm{T}_s\eta_\omega'') > 0 \tag{A.2}$$

From (A.1), (A.2) the condition guaranteeing asymptotic stability (negative semi-definiteness of $\Delta\mathcal{J}_k$) is given as

$$\therefore 0 < \eta_\omega'' < 2\mathrm{T}_s^{-1}$$

$$\Rightarrow 0 < \eta_\omega^k < 2\mathrm{T}_s^{-1}(h_{max}^k)^{-2}$$

$$\therefore 0 < \eta_\omega^k = \frac{\beta_\omega\mathrm{T}_s^{-1}}{(\partial\mathcal{E}_k/\partial\widehat{\omega}_1^k)^2} < \frac{2\mathrm{T}_s^{-1}}{[\partial\mathcal{E}_k/\partial\widehat{\omega}_1^k]_{max}^2} \tag{A.3}$$

where $0 < \beta_\omega < 2$

## Appendix B [30], [34]

**Saturation factor** (SF): In an instrument transformer the ratio of secondary voltage (also called saturation voltage) $\mathbf{V_S}$ to primary voltage (also known as excitation voltage) $\mathbf{V_E}$ is called saturation factor (SF). When SF exceeds unity, it indicates the core is saturated and the waveshape on the secondary side is distorted and rich in harmonics.

**Burden**: Burden of an Instrument system is the rated Volt-Ampere loading (modelled as impedance) which is permissible without errors exceeding the limits for a particular class (protection/ metering) of an instrumentation chain.

## Acknowledgement

The authors acknowledge and thank the authorities of the Power System Operation Corporation (POSOCO), NRLDC India and Prof. B. K. Panigrahi for sharing the field data.

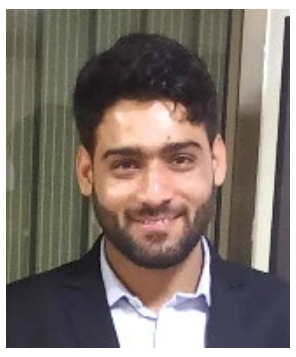

**Abdul Saleem Mir** received his B. Tech. degree (Hons.) in Electrical Engineering from National Institute of Technology, Srinagar, J&K, and Ph.D. degree in Electrical Engineering from Indian Institute of Technology Delhi, India, in 2014 and 2020 respectively. He is currently a Research Fellow at the University of Southampton working on a collaborative project with Imperial College London as collaboration partner. He is a member of IEEE PES Task Force on Dynamic State and Parameter Estimation. His research interests include dynamic state estimation and control, power system dynamics and modeling/control of renewable energy systems.

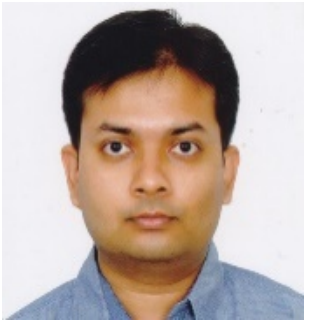

**Abhinav Kumar Singh** (S'12–M'15) received his B. Tech. degree in Electrical Engg. from Indian Institute of Technology Delhi, India, and Ph.D. degree in EE from Imperial College London, U.K. in 2010 and 2015 respectively. He is a Lecturer at the School of Electronics and Computer Science, University of Southampton. Previously, he was a Lecturer at the University of Lincoln from Aug 2017 to Mar 2019, and a Research Associate at Imperial College London from Jan 2015 to July 2017. He won IEEE PES Working Group Recognition Award in 2016 for his contributions to IEEE PES task force on benchmark systems. He is a member of IEEE PES Task Force on Dynamic State and Parameter Estimation. He currently serves as an Editor of IEEE Transactions on Power Systems. His research interests include real-time estimation and control of future energy networks.

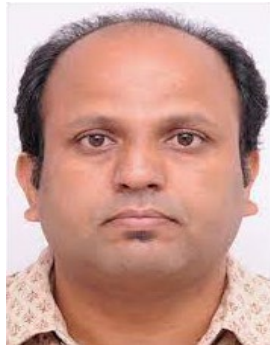

**Nilanjan Senroy** (S'01–M'06-SM'18) received the B. Tech. degree from the National Institute of Technology, Jamshedpur, India, and the M.S. and Ph.D. degrees from Arizona State University, Tempe, AZ, USA. He also has postdoctoral experience at the Center for Advanced Power Systems, Florida State University, Tallahassee, FL, USA. He is currently Power Grid Chair Professor in the Department of Electrical Engineering, Indian Institute of Technology Delhi, India. His research interests include power system stability and control, dynamics, modeling and simulation of wind energy conversion systems and signal processing techniques in power systems.